\documentclass[aps,twocolumn,floatfix,amsmath,amssymb]{revtex4}
\usepackage{graphicx}
\usepackage{dcolumn}
\usepackage{bm}
\usepackage{epstopdf}

\begin{document}
\title{F\"orster interaction induced phase shift in a pair state interferometer}
\author{J. Nipper}
\author{J.B. Balewski}
\author{A.T. Krupp}
\author{S. Hofferberth}
\author{R. L\"{o}w}
\author{T. Pfau}
\affiliation{
5. Physikalisches Institut, Universit\"{a}t Stuttgart, Pfaffenwaldring 57, 70569 Stuttgart, Germany.}

\date{\today}

\begin{abstract}

We present experiments measuring an interaction induced phase shift of Rydberg atoms at Stark tuned F\"orster resonances. The phase shift features a dispersive shape around the resonance, showing that the interaction strength and sign can be tuned coherently. We use a pair state interferometer to measure the phase shift. Although the coupling between pair states is coherent on the time scale of the experiment, a loss of visibility occurs as a pair state interferometer involves three simultaneously interfering paths and only one of them is phase shifted by the mutual interaction. Despite additional dephasing mechanisms a pulsed F\"orster coupling sequence allows to observe coherent dynamics around the F\"orster resonance.
\end{abstract}

\maketitle

Coherent control of strongly interacting gases is of great interest, as they can serve as model systems for correlated quantum many body physics. Especially Rydberg atoms obtain much interest in ultra-cold atomic physics as they offer strong interactions which are tunable both in strength and character. Noteworth applications are for example quantum computing \cite{SWM10} or quantum simulation \cite{WML10} as well as ultracold chemistry \cite{BNB10} and quantum phase transitions \cite{WLP08}.

A promising tool for creation and control of strong interactions are Stark-tuned F\"orster resonances, where the interaction strength and the character, from resonant dipole-dipole interaction to van der Waals interaction, can be tuned by small electric fields \cite{RCK07,AVG98,VVZ06,BPM07,TDN08,RTB10}. Here, we present experiments showing that Rydberg atom pairs close to a F\"orster resonance act coherently, a prerequisite for an interaction induced phase shift used in proposals for Rydberg quantum gates \cite{SWM10,Ja00}. Furthermore we present a direct observation of this tunable phase shift, to the best of our knowledge not observed before. Coherence at F\"orster resonances in Rydberg systems has previously been studied in the coupling between pair states \cite{ARM02} and by means of optical Ramsey spectroscopy in ultracold atomic systems \cite{NBK12}. A decreased coherence time at resonance was measured and quantified based on two-level optical Bloch equations.

To describe the coherent dynamics we observe, we introduce in this paper the concept of a pair state Ramsey interferometer, extending the usual two-level atom interferometer to two interacting two-level atoms coupled to the optical excitation field. Besides the ground state $\left.\left|gg\right>\right.$ and the doubly excited state $\left.\left|rr\right>\right.$ there are two singly excited states $\left.\left|gr\right>\right.$ and $\left.\left|rg\right>\right.$. Only the symmetric combination of the two couples to the Ramsey field. Therefore the pair state version of a Ramsey interferometer consists of three simultaneously interfering paths. Only one of them is affected by the possibly coherent Rydberg-Rydberg interaction $U$ and experiences a phase shift $\varphi(U)$.
\begin{figure}
	\centering
	\resizebox{0.95\columnwidth}{!}{\includegraphics{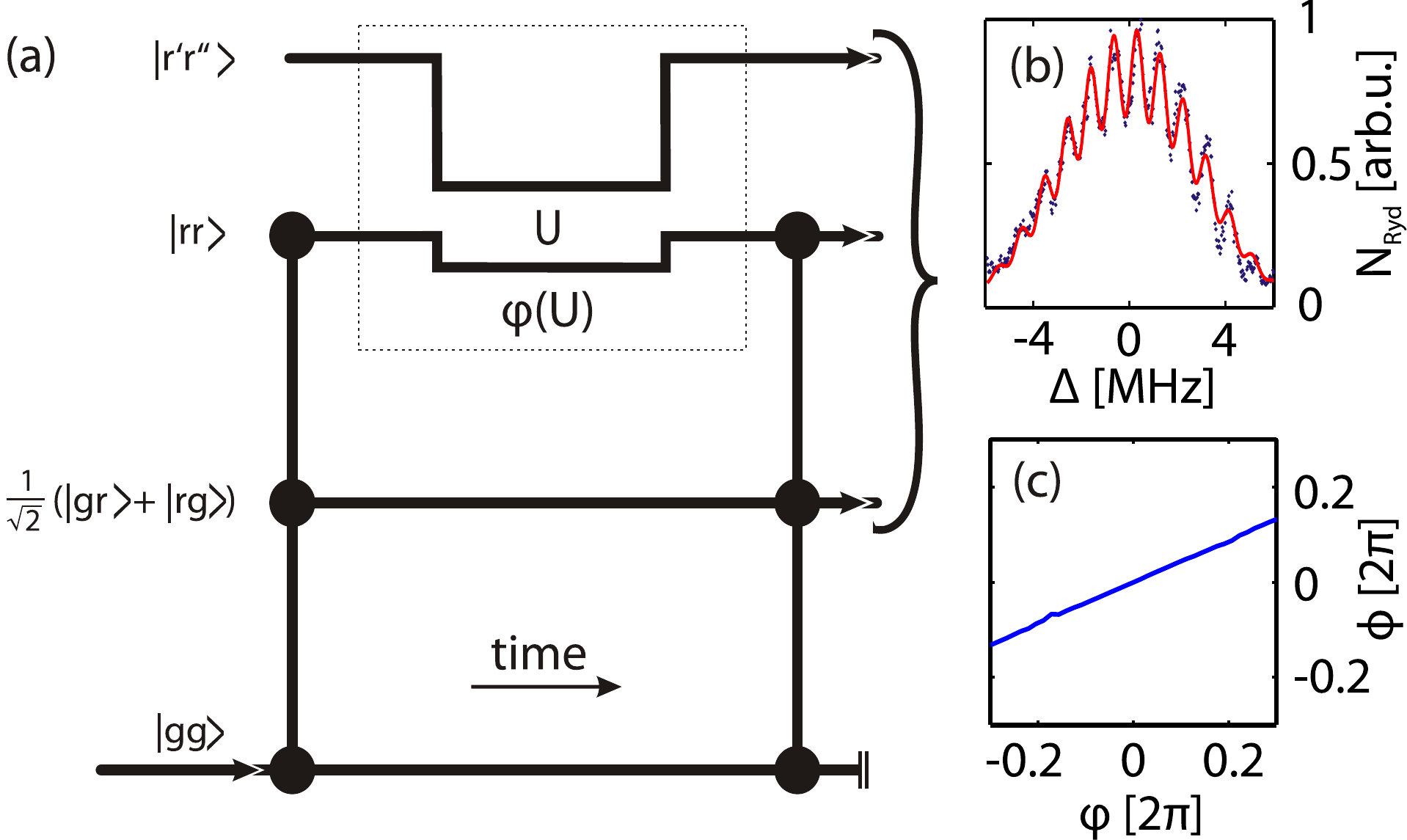}}
	\caption{(a) Schematic of the pair state interferometer. Interactions between Rydberg atoms change the phase $\varphi(U)$ of the $\left.\left|rr\right>\right.$ state relative to the other states and create population in $\left.\left|r^\prime r^{\prime\prime}\right>\right.$. A pulsed Ramsey field couples three states simultaneously. A Rydberg detector detects the number of Rydberg atoms $N_{Ryd}$, resulting in a Ramsey spectrum (b) (blue data\-points) depending on the detuning $\Delta$ of the exciting laser. The red line is a fit to the data. (c) shows the transfer function for 0.6$\pi$-Ramsey pulses describing the dependence of the fitted phase of the Ramsey fringes $\phi$ on the phase shift $\varphi$ in the $\left.\left|rr\right>\right.$ path for small angles.}
	\label{Fig1}
\end{figure}

In Fig. \ref{Fig1}\,(a) a schematic of the relevant pair states for a Ramsey interferometer in an ensemble of atoms with switchable interaction is shown. The interfero\-meter starts in the state where both atoms are in the ground state $\left.\left|gg\right>\right.$. Optical light pulses excite atoms to the Rydberg state and induce a coupling between $\left.\left|gg\right>\right.$, $(\left.\left|gr\right>\right.+\left.\left|rg\right>\right.)/\sqrt{2}$ and $\left.\left|rr\right>\right.$. A switchable interaction can be induced by coupling $\left.\left|rr\right>\right.$ to another Rydberg atom pair state $\left.\left|r^\prime r^{\prime\prime}\right>\right.$. Here, a Stark-tuned F\"orster resonance is used, but the concept of the pair state interferometer is valid for any tunable interaction. In the case of weak interactions (adiabatic regime) a phase shift of $\left.\left|rr\right>\right.$ occurs. Strong interactions (diabatic regime) transfer atoms to $\left.\left|r^\prime r^{\prime\prime}\right>\right.$.
The number of Rydberg atoms, depending on the population of the upper arms, is detected. From a fit to the Ramsey spectrum in frequency space (Fig. \ref{Fig1}\,(b)) the visibility and the phase $\phi$ of the Ramsey fringes can be obtained.

This three path interferometer behaves considerably different than a two path interferometer. For example, even in the case of adiabatically switched interactions and an individual pair of atoms, a coherent phase shift $\varphi(U)$ in the $\left.\left|rr\right>\right.$ path leads to a loss of visibility that can not be avoided. This is one source of the reduced coherence times oberved in Ramsey experiments \cite{NBK12}. Similar effects have also been observed in atom interferometry \cite{MW01}. Additionally the phase shift $\varphi(U)$ translates to a phase shift $\phi$ of the Ramsey fringes. The transfer function from $\varphi(U)$ to $\phi$ is a non-trivial function, depending on the population of the different paths of the interferometer. Fig. \ref{Fig1}\,(c) shows such a simulated transfer function for 0.6$\pi$-Ramsey pulses. For small angles it is a monotonic function in $\varphi(U)$. Close to $\varphi(U)=\pi$ the fringe pattern is strongly disturbed and the phase $\phi$ is not well defined. Under any circumstances the effective phase shift of the Ramsey fringes $\phi$ is smaller than the phase shift of the doubly excited state $\varphi$.

The actual situation in an ensemble with strong binary interactions is more involved. The ensemble average over different interaction strengths due to the distance and angular dependence of the interaction even at constant density results in an additional dephasing. Moreover, an inhomogeneous density distribution will lead to yet another source of dephasing. Furthermore in the experiment decoherence due to a finite excitation linewidth is present. Previous measurements so far can not separate these different sources. 

However, echo or Ramsey type sequences can refocus these inhomogeneous dephasing mechanisms. Ramsey type pulse sequences for the interaction strength allow for the observation of coherent evolution of the ensemble and are studied in this paper close to a F\"orster resonance.

The Stark tuned F\"orster resonances appear if two dipole-dipole coupled pair states are shifted into resonance by a small applied electric field. Here we employ F\"orster resonances in $^{87}$Rb between the pair states $2\cdot44d_{5/2}$ and $46p_{3/2}$\,+\,$42f_{7/2}$, denoted by $\left.\left|dd\right>\right.$ and $\left.\left|pf_i\right>\right.$ respectively. Different magnetic substates of the 42$f$-state lead to several resonances at slighly different electric fields, indicated by the subscript~$i$.
\begin{figure}
\begin{center}
\resizebox{0.95\columnwidth}{!}{\includegraphics{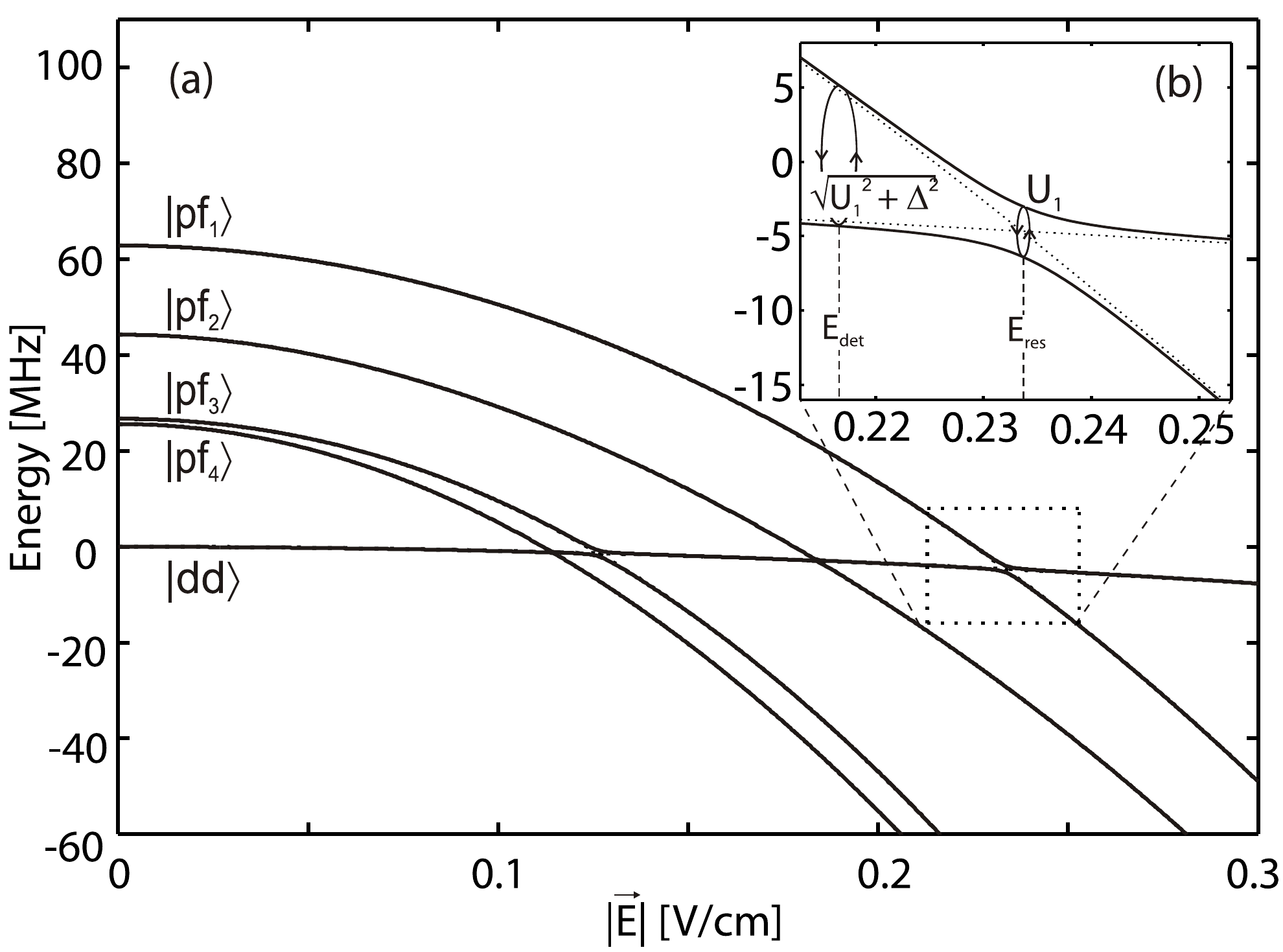}}
\caption{\label{Fig2} (a) Stark map of the relevant pair states (see text). (b) shows a magnification of the Stark map at the electric fields where the pair states are tuned into resonance. The dashed lines denote the pair state energies without coupling, the solid lines include the dipole-dipole coupling for an interatomic distance of 9\,$\mu$m. The energy differences at the resonant electric field $E_{res}$ and at a detuned electric field $E_{det}$ are indicated. }
\end{center}
\end{figure}
Fig \ref{Fig2} shows the Stark shift of the pair states in a 13.55\,G magnetic offset field, parallel to the electric field. The Stark shifts were calculated by diagonalising the single atom Hamiltonian taking the magnetic and electric field into account \cite{NBK12}. In zero electric field a finite F\"orster defect $\Delta=E_{\left.\left|pf_i\right>\right.}-E_{\left.\left|dd\right>\right.}$, the energy difference between the coupled pair states, is present. With increasing electric field the pair states experience different Stark shifts and the F\"orster defect can be tuned (Fig. \ref{Fig2}\,(a)).\\
At a sufficiently large $\Delta$ the interaction between the atoms can be calculated from second order perturbation theory and one obtains a van-der-Waals interaction energy of
\[\Delta E_{vdW}\approx-\frac{|U_i|^2}{\Delta}\]
for every dipole coupled pair state $i$. $U_i$ is the strength of the dipole-dipole coupling
\[U_i(r,\Theta)=\sqrt{2}\cdot\left<\left.pf_i\right|\right.V_{dd}(r,\Theta)\left.\left|dd\right>\right.,\] 
where $V_{dd}=\frac{\vec{p}_1\cdot\vec{p}_2-3(\vec{n}\cdot\vec{p}_1)(\vec{n}\cdot\vec{p}_2)}{r^3}$ is the dipole-dipole operator and $\vec{p}_{1,2}$ are the electric dipole moments of the atoms. The $\sqrt{2}$-factor stems from the degeneracy of $\left.\left|pf_i\right>\right.$ and $\left.\left|f_ip\right>\right.$. This interaction increases with decreasing F\"orster defect and the interaction strength can be tuned by the electric field. For $\Delta=0$ resonant dipole-dipole interaction occurs, resulting in the F\"orster resonance, and the two pair states form an avoided crossing, as shown in Fig. \ref{Fig2}\,(b). For small F\"orster defects $\Delta\approx U_i$ the eigenstates have to be obtained by diagonalization.

The strength of the dipole-dipole coupling $U_i(r,\Theta)$ is calculated as in \cite{RCK07,NBK12}. The experiments are performed in an extended sample, thus the angular dependency is averaged out. For this paper, interactions at finite F\"orster defect $\Delta$ are most relevant where the interaction is dominated by van-der-Waals interaction. Therefore the quadratic mean of the angular dependent interaction is calculated. This gives values of $U_1(r)$\,=\,719\,MHz$\cdot \mu$m$^3/r^3$, $U_2(r)$\,=\,200\,MHz$\cdot \mu$m$^3/r^3$, $U_3(r)$\,=\,654\,MHz$\cdot \mu$m$^3/r^3$ and $U_4(r)$\,=\,167\,MHz$\cdot \mu$m$^3/r^3$ for the resonances at 0.23\,V/cm, 0.18\,V/cm, 0.13\,V/cm and 0.12\,V/cm respectively.

The experiments are performed in a magnetically trapped and evaporatively cooled cloud in the f\,=\,2, $m_f$\,=\,2 state. After cooling the magnetic offset field is ramped to 13.55\,G. This results in an atomic cloud of about 700\,nK temperature at a peak density of about $2\cdot10^{12}$/cm$^3$. In this high offset field the cloud extends to 1/e-radii of 114\,$\mu$m longitudinally and 22\,$\mu$m radially. Further information about the experimental setup can be found in \cite{R12}.

Rydberg atoms are excited by two-photon excitation via the 5p$_{3/2}$-state, blue detuned by $2\pi\cdot400$\,MHz to the intermediate state to preserve the coherence in the excitation process. The total laser linewidth of the two-photon transition is below $2\pi\cdot100$\,kHz and the single atom Rabi frequency is about $\Omega_0$\,=\,$2\pi\cdot25$\,kHz. Optical Ramsey spectroscopy is realized throughout this paper by two short laser pulses of $\tau_p$\,=\,150\,ns duration, separated by a delay time of $\tau_{del}$\,=\,800\,ns  as shown in Fig. \ref{Fig2}\,(c). The pulse area is small enough that the system is not driven into saturation, but a collective enhancement of the excitation occurs \cite{HRB07}. After this pulse sequence the Rydberg atoms are field ionized and detected in an ion detector. As the field ionization is not state selective the total Rydberg atom number $N_{Ryd}$ is measured, independent of the Rydberg states the atoms populate. Within one atomic sample the sequence of excitation and detection is repeated 401 times. Thereby a whole spectrum ranging from -6\,MHz to +6\,MHz around the atomic resonance is measured in one atomic sample without the need of averaging over different samples. As in Fig. \ref{Fig1}\,(b), these Ramsey spectra show typical Ramsey fringes, which can be fitted to obtain the visibility $V$
and the phase $\phi$. The visibility in a pair state interferometer is affected by a pair interaction phase shift $\varphi(U)$, a population transfer between the Rydberg pair states and by dephasing and decoherence processes. The phase of the Ramsey fringes $\phi$ provides information about $\varphi(U)$ according to the transfer function.

Between the optical Ramsey pulses an electric field can be tuned, enhancing the interaction close to a F\"orster resonance during the delay time only. Thereby interaction effects like Rydberg blockade induced saturation during the excitation pulses can be strongly diminished.

\begin{figure}[t]
\begin{center}
\resizebox{0.95\columnwidth}{!}{\includegraphics{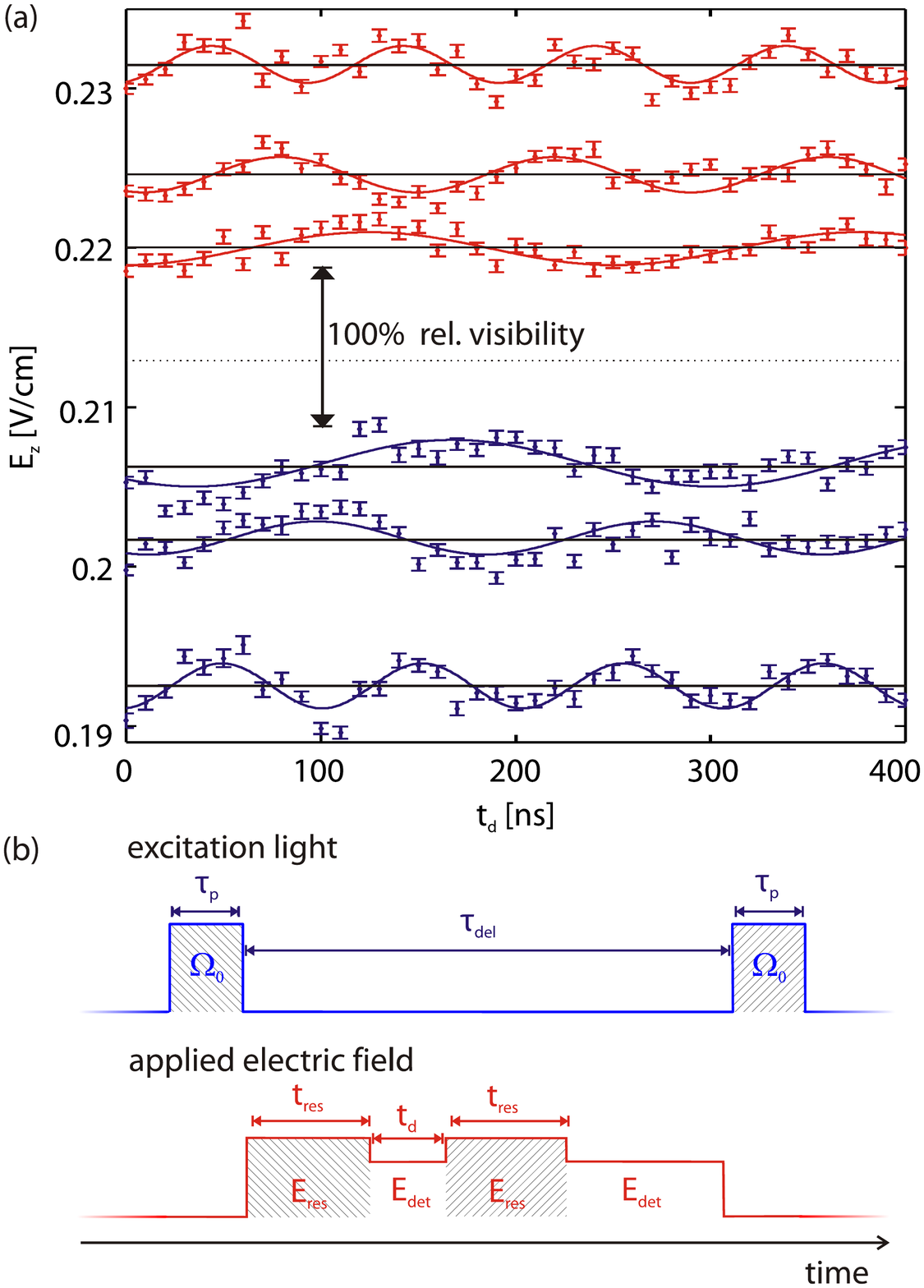}}
\caption{\label{Fig3} (a) Oscillations in the visibility are measured for the pulse sequence (b) of the double Ramsey experiment. The electric field during the delay time $E_{\text{det}}$ is indicated by the solid black lines. The oscillations in the visibility (data points with errorbars resulting from the standard deviation of the fit to the Ramsey spectrum) are centered around the applied electric field. The solid lines are sinusoidal fits to the data. Blue (red) data indicate positive (negative) F\"orster defects. The dotted line indicates the position of the F\"orster resonance.}
\end{center}
\end{figure}

In an experiment close to the F\"orster resonance the $\left.\left|pf\right>\right.$ state couples only to $\left.\left|dd\right>\right.$ via electric field dependent dipole-dipole coupling, but it does not couple to the light field, realizing an interferometer as depicted in Fig. \ref{Fig1}. This offers the possibility to study the coherent evolution of the subsystem $\left.\left|dd\right>\right.$, $\left.\left|pf\right>\right.$ (dashed box in Fig. \ref{Fig1}) separately by applying a Ramsey-like electric field sequence, similar to \cite{ARM02}. The pulse sequence for this double-Ramsey experiment can be found in Fig. \ref{Fig3}\,(b) and is comparable to Ramsey experiments on Fesh\-bach resonances \cite{D02}, where a similar magnetic field sequence was used. Here, the electric field is first pulsed for $t_{\text{res}}$\,=\,200\,ns to the electric field $E_{\text{res}}$ (see Fig. \ref{Fig2}\,(b)) to tune the pair states into resonance. The rise time of the electric field is about 20\,ns. After a variable delay time $t_{\text{d}}$ between 0\,ns and 400\,ns a second 200\,ns pulse at $E_{\text{res}}$ is applied. Between and after these two Ramsey-like pulses the electric field is detuned from exact resonance to a variable value $E_{\text{det}}$.

This experiment can be regarded as a Ramsey-like experiment between the $\left.\left|dd\right>\right.$ and $\left.\left|pf\right>\right.$ states only. The first resonant electric field pulse couples the $\left.\left|dd\right>\right.$-state to the $\left.\left|pf\right>\right.$-state, generating a coherence between these states. During the delay time $t_{\text{d}}$ the electric field is detuned from exact resonance (Fig. \ref{Fig2}\,(b)). The atom pairs will oscillate between $\left.\left|dd\right>\right.$ and $\left.\left|pf\right>\right.$ with the oscillation frequency $\sqrt{U^2+\Delta^2}$. The second resonant electric field pulse interferes both pair states again. 
Depending on the F\"orster defect at $E_{\text{det}}$ and on the delay time $t_{\text{d}}$ the atom pairs can be refocused on $\left.\left|dd\right>\right.$ and a high visibility occurs in the optical Ramsey spectrum.
The oscillations between the pair states are now visible as oscillations in the visibility of the Ramsey fringes when the delay time $t_\text{d}$ is varied.

\begin{figure}[t]
\begin{center}
\resizebox{0.95\columnwidth}{!}{\includegraphics{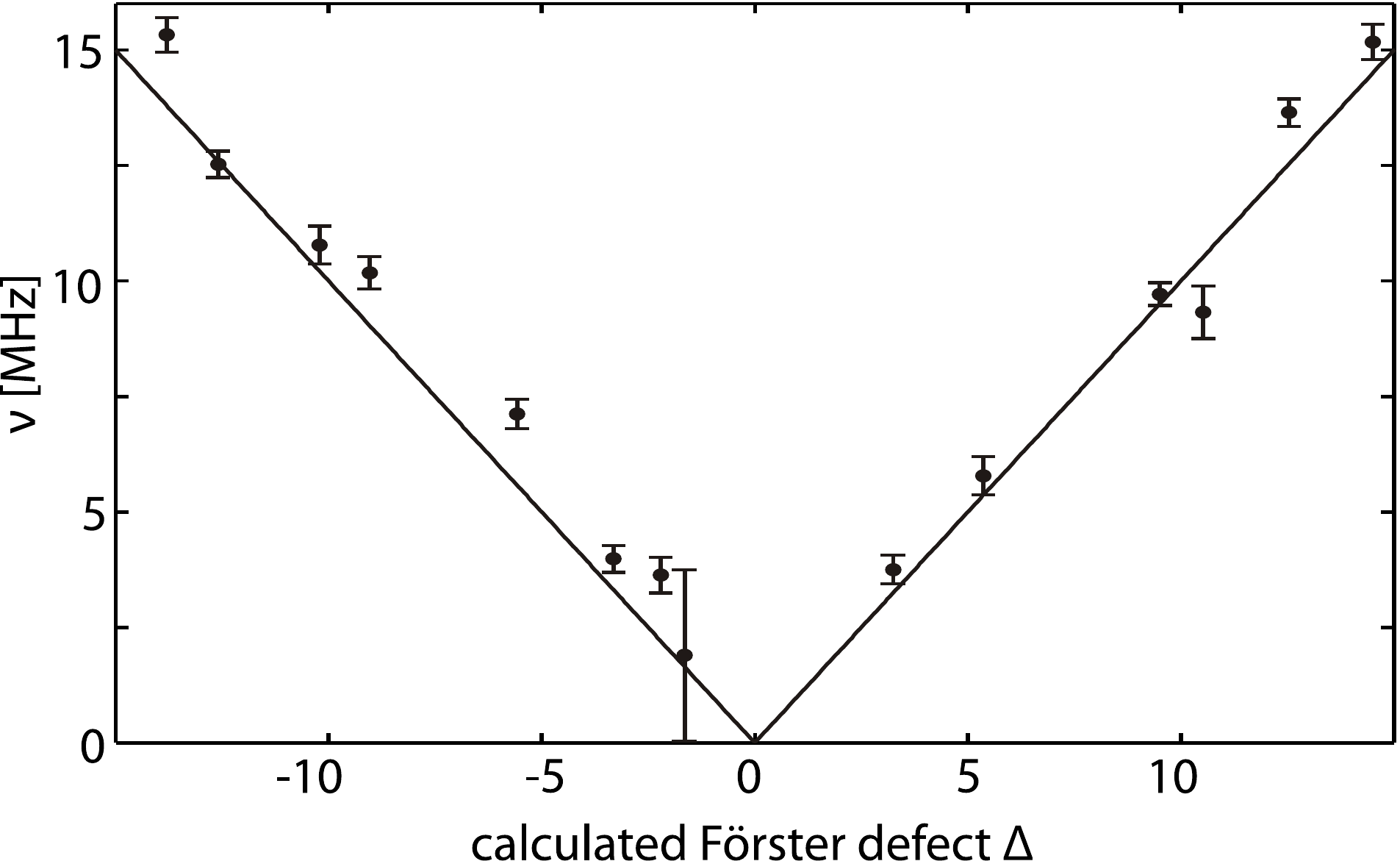}}
\caption{\label{Fig4} Measured frequency $\nu$ of the oscillation in the visibility (datapoints) versus the calculated F\"orster defect. The errors are the standard deviation of the sinusoidal fits. The solid line shows $\nu$\,=\,$|\Delta|$.}
\end{center}
\end{figure}

This particular sequence offers the advantage that the on- and off-resonant electric field pulse lengths are constant. Thereby additional losses of coherence, e.g. due to inhomogeneous atom distributions, reduce the visibility to a constant value throughout this experiment. Oscillations in the visibility when varying the delay time between the resonant pulses can clearly be separated. Fig. \ref{Fig3}\,(a) shows a selection of such double-Ramsey experiments for different values of $E_{\text{det}}$, indicated by the solid black lines. Oscillations in the visibility are clearly visible and no damping can be observed, indicating that the two-body coupling between the pair states is coherent at least on the timescale of the experiments. The oscillation frequency is obtained from sinusoidal fits to the data. It shows a minimum at the position of the F\"orster resonance at $E_z$\,=\,0.213\,V/cm \cite{NBK12}, as expected. The slight mismatch to the theory presented in Fig. \ref{Fig2} results from an unknown radial electric offset field $E_r$ on the order of 0.05\,V/cm and a slight misalignment between the applied electric and magnetic field, defining the quantisation axis, that can not be included in the calibrated component $E_z$ of the electric field. For the resonance here $E_z\gg E_r$ is valid, therefore we can approximately compensate for the mismatch by a constant offset of 0.018\,V/cm in the electric field. Taking this offset into account the F\"orster defect can be calculated, given the calculated Stark shifts in Fig. \ref{Fig2}. In Fig. \ref{Fig4} the measured oscillation frequency versus the calculated F\"orster defect is plotted. The oscillation frequency follows the F\"orster defect $\Delta$, as expected for $\Delta\gg U$. Close to the F\"orster resonance a deviation from the linear behaviour is expected. However, there the amplitude of the oscillations is strongly reduced and no signal could be obtained for F\"orster defects of $|\Delta|\lesssim 2$\,MHz ($E_z\approx0.209\ldots0.217$\,V/cm).

This is also true for a slightly different pulse sequence where the electric field is tuned to zero after the second Ramsey-like electric field pulse. This sequence converges to a single-pulse experiment for $E_{\text{det}}\rightarrow E_{\text{res}}$, where only the total pulse length is varied. Instead of direct Rabi-oscillations only an exponential loss of visibility could be observed, as predicted in \cite{RTB10a}.

Interaction strengths in the range of 2\,MHz are expected at interatomic distances of 7\,$\mu$m, giving a lower limit on the average Rydberg atom distance. This is a reasonable value as the Rydberg signal is obtained from the whole cloud, including the wings, averaging the interatomic distance to rather large values. We interprete this experiment as follows: Strong dephasing occurs when the evolution of the system is dominated by the interaction between the pair states, as this interaction energy forms a broad band due to the dependence on the interatomic distance \cite{CP10}. However, the coupling between single pair states is coherent on the timescale of the experiment. This coherence leads to the observed oscillations in the interferometer if the time evolution is dominated by the spatially constant F\"orster defect.

\begin{figure*}
	\centering
		\includegraphics[width=0.85\textwidth]{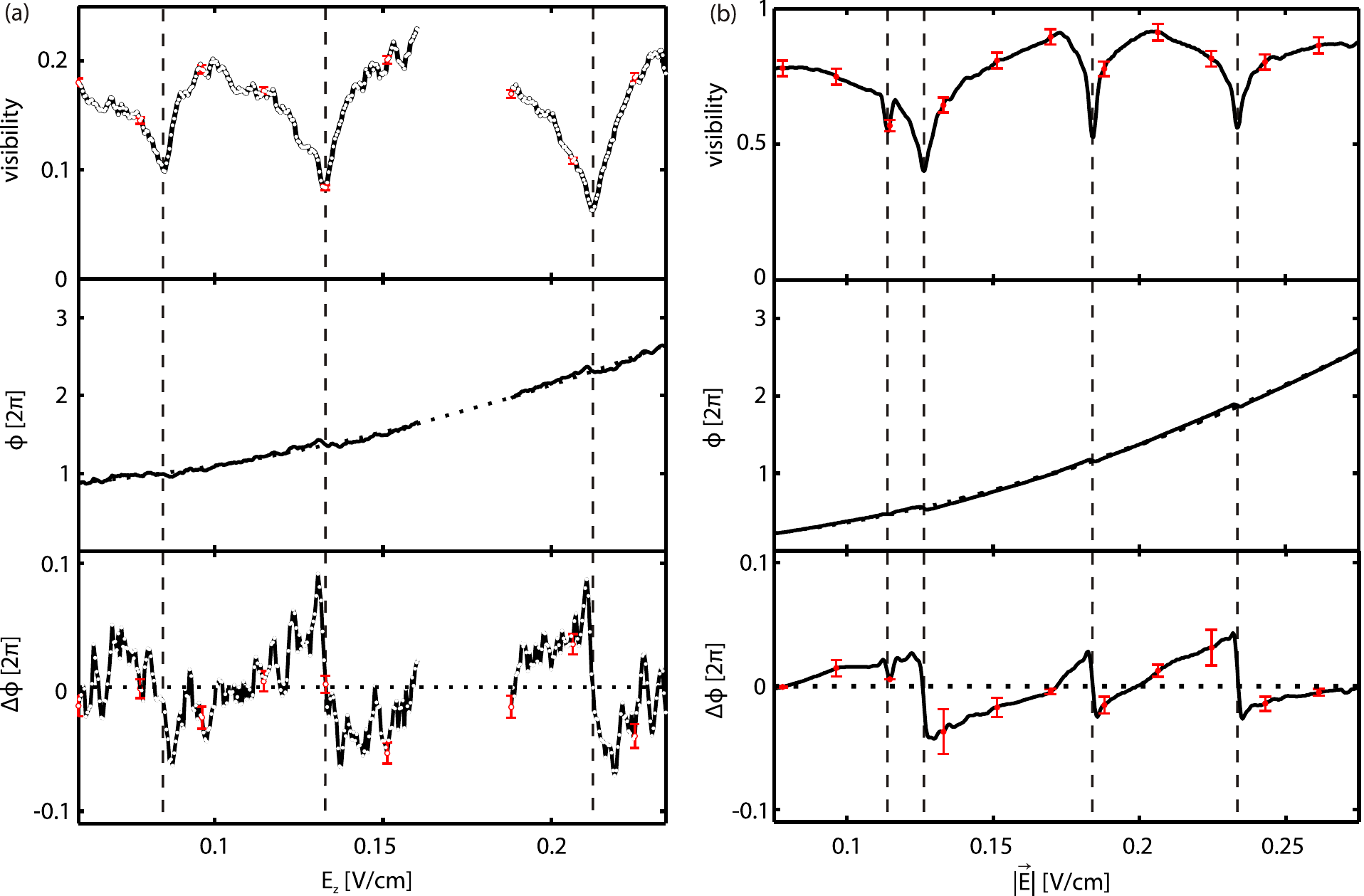}
	\caption{Visbility and phase obtained from fits to the measured (a) and simulated (b) Ramsey spectra versus the electric field. Note that the experimental spectra are plotted versus the calibrated component of the electric field $E_z$ and the simulated data versus the total electric field $|\vec E|$. The uppermost panels show the visibility, the middle panel shows the phase of the Ramsey fringes (solid line) and a quadratic fit to the data (dotted line) and the lower panels show the difference of the measured phase to the pure quadratic behavior. Exemplarily some errorbars are plotted, denoting the standard deviation of the fit parameter.}
	\label{Fig5}
\end{figure*}

Despite the fact that this coupling is coherent, it leads to a loss of visibility in the three path pair state interferometer. However, the phase shift $\varphi$ translates also in a measurable but small phase shift $\phi$ in the Ramsey fringe pattern. To study this a single electric field pulse during the whole length of the delay time between the optical Ramsey pulses was used. The visibility and the phase of the measured Ramsey spectrum are obtained for varying strength $E_z$ of the pulsed electric field. Fig. \ref{Fig5}\,(a) shows the results of a fit to the measured Ramsey spectra. At the positions of the F\"orster resonances the visibility is reduced because of a population in $\left.\left|pf\right>\right.$ and a phase shift $\varphi$ of $\left.\left|dd\right>\right.$. Three distinct dips can be seen due to a substructure of the F\"orster resonance \cite{NBK12}. The phase shows a quadratic dependence as it is shifted by the quadratic Stark effect of the Rydberg atoms. If this pure quadratic effect is subtracted a deviation $\Delta\phi$ is visible that shows a clear dispersive behavior centered around the positions of the F\"orster resonances.

This interaction induced phase shift can be best understood in the interferometer picture of Fig. \ref{Fig1}. Here, the three states coupled by the light field are $\left.\left|gg\right>\right.$, $\frac{1}{\sqrt 2}(\left.\left|gd\right>\right.+\left.\left|dg\right>\right.)$ and $\left.\left|dd\right>\right.$. A coupling between $\left.\left|dd\right>\right.$ and $\left.\left|pf\right>\right.$ leads to an interaction induced phase shift of $\left.\left|dd\right>\right.$ during the delay time, when the system is tuned close to the F\"orster resonance. 
For a first estimate, taking an interaction strength of 1\,MHz for two Rydberg atoms at F\"orster resonance and an interaction time of 0.8\,$\mu$s, one would expect a phase shift of the pair state of $\varphi$\,=\,$0.8\pi$. The transfer function of the three path interferometer (see Fig. \ref{Fig1}) lowers the observed shift $\phi$. As the switching of the electric fields in the experiment is not adiabatic, a further reduction of $\phi$ is expected. The sign of the phase shift is determined by the direction of the interaction induced energy shift of $\left.\left|dd\right>\right.$ which changes sign at the resonance position and a dispersive phase effect occurs.

This phase effect not only directly verifies the coherence of the interaction but also shows that the strength and the sign of the interaction can be tuned by the electric field. The interaction switches from attractive at electric fields smaller than the resonant field $E_{\text{res}}$ to repulsive above the resonance. Exactly on resonance two equally spaced states in the avoided crossing (Fig. \ref{Fig2}) above and below the unperturbed states appear. Under these conditions the system is diabatically switched from the unperturbed $\left.\left|dd\right>\right.$ state at $E$\,=\,0\,V/cm to the perturbed states at $E$\,=\,$E_{\text{res}}$, generating a superposition state that does not experience interactions and no phase shift occurs.

To model this experiment the Schr\"odinger equation is solved numerically for the experimental sequence. Taking only one F\"orster resonance into account the Hamiltonian can be expressed in the basis ($\left.\left|gg\right>\right.$,$\frac{1}{\sqrt{2}}(\left.\left|gd\right>\right.+\left.\left|dg\right>\right.)$,$\left.\left|dd\right>\right.$,$\frac{1}{\sqrt{2}}(\left.\left|pf\right>\right.+\left.\left|fp\right>\right.)$) as
\[H=	
\begin{pmatrix}
0 & \frac{\Omega}{\sqrt{2}} & 0 & 0 \\
 \frac{\Omega}{\sqrt{2}} & \delta_{L}+E_{\left.\left|d\right>\right.} & \frac{\Omega}{\sqrt{2}} & 0\\
 0 & \frac{\Omega}{\sqrt{2}} &   2\delta_{L}+2E_{\left.\left|d\right>\right.} & U(r)  \\
 0 & 0 & U(r) & 2\delta_L+E_{\left.\left|pf\right>\right.}\\
 \end{pmatrix}.
 \]
The antisymmetric pair states are not coupled. $E_{\left.\left|d\right>\right.}$ is the Stark shift of one atom in the 44d-state, $E_{\left.\left|pf\right>\right.}$ the shift of the $\left.\left|pf\right>\right.$-pair state and $\delta_L$ the detuning of the laser to the 44d-state. The Stark shifts are obtained from the calculations in Fig. \ref{Fig2}\,(a). The extension to four resonances, described by a seven-dimensional Hamiltonian, is straightforward. 

To account for the Rydberg atom distribution the calculated Ramsey spectra for several radii $r$ were weighted and averaged according to a Chandrasekhar distribution \cite{C43}
\[P(r)=e^{-r^3/r_0^3}3r^2/r_0^3,\]
describing the nearest-neighbor distribution at the average distance $r_0$. These spectra were fitted as in the experiment to extract the visibility and phase. Best results were obtained for an average distance of 9\,$\mu$m, in good agreement with the observation from the double-Ramsey experiment, and for a Rabi frequency of $\Omega$\,=\,2\,MHz. This frequency is enhanced relative to the single atom Rabi frequency by the collective excitation process. It is on the order of what is expected from simple estimates of the number of atoms per blockade sphere for a 44d C$_6$-coefficient of 27\,GHz$\cdot\mu$m$^6$ at $|\vec E|$\,=\,0\,V/cm.

This model simplifies the actual system in many ways. It neglects the angular dependence of the interaction, only binary, next-neighbor interactions are calculated and the many-body nature of the experiment is considered solely by a collective enhancement of the Rabi frequency. Nevertheless the model effectively allows to understand the fundamental aspects of the experimental findings.

Fig. \ref{Fig5}\,(b) shows the calculated visibility and phase versus the total electric field $|\vec E|$. The absolute position of the measured resonances are shifted relative to the calculations by uncontrolled radial electric fields. The visibility in the experiment is roughly a factor 4 smaller than in the calculations and the measured dips are broader. This mismatch can be explained by additional decoherence processes beyond the pair state interferometer. The minimal linewidth of the transition to the 44d-state at 700\,nK temperature was measured to be 300\,kHz, broadened by the magnetic field gradient in the trap and possibly by electric field inhomogeneities. This considerably reduces the visibility on the 1\,$\mu$s time scale of the experiment and is not included in the calculations. However, qualitatively the observed visibility at the F\"orster resonances can be reproduced with the two-body calculation as described above.

The phase, on the other hand, is not expected to be substantially disturbed by an additional loss of coherence and agrees remarkably good with the calculation. The dispersive shape of the signal and the amplitude are reproduced. Nevertheless, the phase of the Ramsey fringes is a nontrivial function of the populations of the pair states and the interaction strength. Thereby it strongly depends on the Rabi frequency and on the Rydberg atom distribution. In a saturated ensemble blockade effects clearly affect the nearest neighbor distribution and many-body effects \cite{MCT98,YRP09} beyond the two-body calculation occur. The inhomogeneous density in a trapped cloud will alter the distribution as well. This might account for the slight mismatch in Fig. \ref{Fig5}.

In conclusion, we demonstrated the coherent coupling between pair states at a F\"orster resonance for Rydberg atoms and we observed an interaction induced phase shift on the atoms. The dispersive shape of the phase shift shows the tunability of the strength and the sign of the interaction. Fully coherent simulations of the introduced pair state interferometer reproduce the observed phase shift and the loss in visibility.

In single atom experiments \cite{JUH08} individual single site adressing $\pi$-excitation pulses can be used and the system can be reduced to a two path interferometer. There phase shifts on the order of $\pi$, as necessary for applications like phase gates, are realistic. Therefore we see these results as a step towards controlled phase gates \cite{PRS02} and quantum simulation, e.g. of energy transport processes \cite{PH08} in quantum networks.

Due to the strong distance dependence of the interaction, F\"orster resonances can be used as a spectroscopic ruler \cite{SH67}. This might offer a tool to gain more insight into the Rydberg atom correlation function, which under certain conditions is expected to show a crystalline order \cite{WLP08}.
\\
\begin{acknowledgments}
We thank I.I. Ryabtsev for valuable discussions. This work is funded by the Deutsche Forschungsgemeinschaft (DFG) within the SFB/TRR21 and the project PF~381/4-2. We also acknowledge support by the ERC under contract number 267100.
\end{acknowledgments}

\end{document}